\documentclass[3p,12pt,numbers, preprint]{elsarticle}
\usepackage[utf8]{inputenc}
\makeatletter
\def\ps@pprintTitle{%
    \let\@oddhead\@empty
    \let\@evenhead\@empty
    \def\@oddfoot{}%
    \let\@evenfoot\@oddfoot

}
\makeatother
\usepackage{amsmath}
\usepackage{amsfonts}
\usepackage{amssymb}
\usepackage{bm}

\usepackage{geometry}

\biboptions{sort&compress}
\usepackage[
]{hyperref}

\graphicspath{ {./figs/} }

\begin{document}
\begin{frontmatter}
\title{Collective Invasion: When does domain curvature matter?}

\author[1,2]{J. J. Pollacco\corref{cor1}}
\ead{joseph.pollacco@bioch.ox.ac.uk}
\author[2]{R. E. Baker\corref{contrib}}
\ead{ruth.baker@maths.ox.ac.uk}
\author[2]{P. K. Maini\corref{contrib}}
\ead{philip.maini@maths.ox.ac.uk}

\affiliation[1]{organization={Department of Biochemistry},
addressline={University of Oxford},
postcode={OX1 3QU},
city={Oxford},
country={United Kingdom}}

\affiliation[2]{organization={Wolfson Centre for Mathematical Biology, Mathematical Institute},
addressline={University of Oxford},
city={\\ Oxford},
postcode={OX2 6GG},
country={United Kingdom}}

\cortext[contrib]{Authors contributed equally}
\cortext[cor1]{Corresponding author}

\begin{abstract}
Real-world cellular invasion processes often take place in curved geometries. Such problems are frequently simplified in models to neglect the curved geometry in favour of computational simplicity, yet doing so risks inaccuracy in any model-based predictions. To quantify the conditions under which neglecting a curved geometry are justifiable, we examined solutions to the Fisher-Kolmogorov–Petrovsky–Piskunov (Fisher-KPP) model, a paradigm nonlinear reaction-diffusion equation typically used to model spatial invasion, on an annular geometry. Defining $\epsilon$ as the ratio of the annulus thickness $\delta$ and radius $r_0$ we derive, through an asymptotic expansion, the conditions under which it is appropriate to ignore the domain curvature, a result that generalises to other reaction-diffusion equations with constant diffusion coefficient. We further characterise the nature of the solutions through numerical simulation for different $r_0$ and $\delta$. Thus, we quantify the size of the deviation from an analogous simulation on the rectangle, and how this deviation changes across the width of the annulus. Our results grant insight into when it is appropriate to neglect the domain curvature in studying travelling wave behaviour in
reaction-diffusion equations.
\end{abstract}
\end{frontmatter}

\section{Introduction}
Reaction-diffusion models are frequently used in applied mathematics to model invasion processes, finding use in collective cell migration and wound healing \cite{Arciero2021, Sherratt1990, Maini2020, Maini2004}, tumour growth \cite{Yin2019}, and in ecology \cite{SKELLAM1951}. Invasion processes on curved domains are prolific in nature, and so an emerging trend in both experiment and modelling of such processes is to examine invasion in curved geometries  \cite{Treloar2014,Jin2018,Simpson2013,Buenzli2022}. However, it is often desirable to simplify a calculation or simulation by neglecting the curvature of the domain, for example in modelling of the neural crest, a powerful paradigm of collective cell migration \cite{giniunaite_modelling_2020}. Therefore, understanding the impact of domain curvature in such models is critical to ensure the model predictions are accurate. The starting point for many reaction-diffusion models in two spatial dimensions is the general reaction-diffusion equation for the scalar field $u(x, y, t)$
\begin{equation}
    \label{eqn:generalrxndiff}
    \frac{\partial u}{\partial t} = \nabla \cdot (D(u) \nabla u) + F(u),
\end{equation}
where $D(u) > 0$ for $\forall u$ and $F(u)$ is a function such that Eq. \eqref{eqn:generalrxndiff} has at least one positive stable steady state in the absence of diffusion. We focus our attention on the paradigm reaction-diffusion equation for studying spatial invasion, the single-species Fisher-KPP equation \cite{FISHER1937,KPP1937}. The Fisher-KPP equation is given by
\begin{equation}
    \label{eqn:simplestfisher}
    \frac{\partial u}{\partial t} = D \nabla^2 u + ku \left (1-\frac{u}{K} \right ),
\end{equation}
with $D$, $k$ and $K$ positive constants. The Fisher-KPP equation is well-studied in one dimension and permits travelling wave solutions with speed $c \geq 2\sqrt{Dk}$ \cite{Murray2002}, with the equality achieved as $t \to \infty$ when compactly supported initial data are used \cite{aronson1975nonlinear,Aronson1978}.  To gain insight, we study the Fisher-KPP equation on an annular domain, exploring the deviation of the solutions from those simulated on rectangles.

\section{Results: The Fisher-KPP equation on an annulus}
\subsection{Asymptotic expansion of the Fisher-KPP equation in an annular geometry.}
\label{sec:asymptotics}
\begin{figure}[t]
    \centering
    \includegraphics[scale=1]{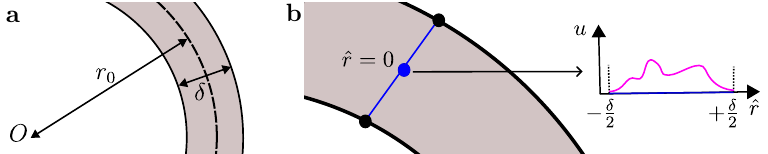}
    \caption{\textbf{a} Definition of the annulus thickness $\delta$ and its radius $r_0$. \textbf{b} Demonstration of the solution case we consider in Sec. \ref{sec:resultsfisheranalysis}. An arbitrary non-zero solution profile $u(r, \theta, t)$ (magenta) is considered on the annulus on a line of constant $\theta$ (blue) at one time. The solution profile along that line of constant $\theta$ is further represented as a function of $\hat{r}$ to the right of the annulus.}
    \label{fig:analysisdiagram}
\end{figure}

For an annular geometry, we define its radius $r_0$ as the radius of the circle defining all points equidistant from its inner and outer circles, and the thickness $\delta$ as the difference between the radii of the outer and inner circles (Fig. \ref{fig:analysisdiagram}\textbf{a}). We define the parameter $\epsilon = \delta / r_0$. In this section we argue that, on the annulus, domain curvature does not substantially affect the solution profile obtained when $\epsilon$ is small. We further show that radial dynamics dominate the solution, and that the azimuthal dynamics provide only a subleading correction of $\mathcal{O}(\epsilon^2)$.
\label{sec:resultsfisheranalysis}
We work in a polar coordinate system $\hat{r} = r - r_0$ with  $\hat{r} \in [-\delta /2, \delta /2]$ and $\theta$ the usual azimuthal angle. The Fisher-KPP equation for $\hat{u}(\hat{r}, \theta, t)$ reads
\begin{equation}
\label{eqn:fisher-hats}
    \frac{\partial \hat{u}}{\partial t} = D \left ( \hat{u}_{\hat{r} \hat{r}} + \frac{\hat{u}_{\hat{r}}}{r_0 + \hat{r}} + \frac{\hat{u}_{\theta \theta}}{(r_0 + \hat{r})^2}\right ) + k\hat{u} \left (1-\frac{\hat{u}}{K} \right ).
\end{equation}
Since we are interested in determining the relative contributions of the radial and azimuthal terms in the Laplacian in Eq. \eqref{eqn:fisher-hats}, we rescale to choose the timescale of interest, $T$, to be the characteristic time for diffusion across the thickness of the annulus. We also rescale the radial coordinate on this length scale, and normalise $\hat{u}$, giving
\begin{align*}
    T &= \delta^2 D^{-1}, & \bar{k} &= k \delta^2 D^{-1}, & \hat{r} &= \rho \delta, & t &= T \tau, & \hat{u} &= K u.
\end{align*}
Writing $\hat{u}(\hat{r}, \theta, t) / K = u(\rho, \theta, \tau)$, we obtain
\begin{equation}
\label{eqn:fisher-beforeexpansion}
    \frac{\partial u}{\partial \tau} = u_{\rho \rho} + \frac{1}{\rho + \frac{r_0}{\delta}} u_{\rho} + \frac{1}{(\rho + \frac{r_0}{\delta})^2} u_{\theta \theta} + \bar{k} u(1-u).
\end{equation}
Rewriting Eq. \eqref{eqn:fisher-beforeexpansion} in terms of $\epsilon$ gives
\begin{equation}
    \frac{\partial u}{\partial \tau} = u_{\rho \rho} + \frac{\epsilon}{1 + \epsilon \rho} u_{\rho} + \frac{\epsilon^2}{(1 + \epsilon \rho)^2} u_{\theta \theta} + \bar{k} u(1-u).
\end{equation}
We assume there is some variation in $u$ in the radial direction (Fig. \ref{fig:analysisdiagram}\textbf{b}), which to be observable must be over a length scale $\sim \delta $, so that the $\rho$ gradients are non-zero. Assuming $\epsilon$ to be small, we further asymptotically expand $u(\rho, \theta, \tau) = u_0(\rho, \theta, \tau) + \epsilon u_1(\rho, \theta, \tau) + \epsilon^2 u_2(\rho, \theta, \tau) + \ldots$. The dynamics of $u_0$ are determined by
\begin{equation}
\label{eqn:generalzerothorder}
    u_{0, \tau} - u_{0, \rho \rho} - \bar{k}u_0 (1-u_0) = 0.
\end{equation}
This is a Fisher-KPP equation in one dimension, with spatial variation in the radial coordinate $\rho$. Thus the dynamics of $u_0$, in the radial direction, are the leading order contribution to the solution. It is easy to show that the dynamics of $u_1$ are determined again only by terms which depend on gradients in $\rho$. $u_0$ and $u_1$ then influence higher order terms, whose temporal evolution is determined by a combination of azimuthal and radial gradients at $\mathcal{O}(\epsilon^2)$. Thus, the dominant variability of the solution is in the radial direction to $\mathcal{O}(\epsilon)$. This conclusion generalises to arbitrary autonomous reaction terms, since it is not required to specify $F(u)$ to arrive at analogous results including only radial gradients at $\mathcal{O}(1)$ and $\mathcal{O}(\epsilon)$. For the Fisher-KPP equation, the existence of a stable steady state allows the dynamics in the azimuthal direction to eventually dominate. The following sections explore numerically the consequences of this result for differing $\epsilon$.

\subsection{Numerical simulation demonstrates similarity to solutions on the rectangle and a timescale separation between radial and azimuthal dynamics.}
\label{sec:initial-numerics}
\begin{figure}[t]
    \centering
    \includegraphics{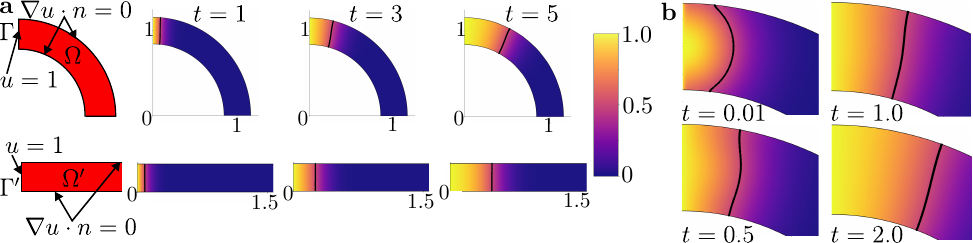}
    \caption{Solution profiles for the Fisher-KPP equation simulated in two dimensions on the rectangle and annulus. Only the upper half of the half-annulus is shown for simplicity. \textbf{a} Demonstration of the simulated geometry. Solutions with a Dirichlet boundary condition $u=1$ on the boundaries $\Gamma, \Gamma'$, and no-flux boundary conditions on the others, evaluated at $t=1, 3, 5$. The $u(x_i, y_i, t)= 0.5$ isoline, found by selecting coordinates $(x_i, y_i)$ with $u$ satisfying $u \in \{|u-0.5|\leq 0.005 \}$, is indicated by a black line. \textbf{b} Solutions with two-dimensional Gaussian initial condition and Dirichlet boundary condition evaluated at $t=0.01, 0.50, 1.00$ and $2.00$.}
    \label{fig:solution_profiles}
\end{figure}

Consider a half-annular domain $\Omega$, with radius $r_0$ and thickness $\delta$ as above, spanning $\theta \in [\pi / 2, -\pi / 2]$. Taking $\delta=0.316$ and $r_0=1$ to set $\epsilon=0.316$, we simulated Eq. \eqref{eqn:simplestfisher} on $\Omega$ using the finite element method implemented in FENICS \cite{AlnaesEtal2015,LoggEtal2012,LoggWells2010,LoggEtal_10_2012,LoggEtal_11_2012,OlgaardWells2010,Kirby2004,kirby2010,alnaes2010}, choosing $K=1$, $D=0.005$, $k=1$ in appropriate units. In the one-dimensional infinite spatial domain case, the width of the wavefront depends on $(\sqrt{Dk})^{-1} $ \cite{Murray2002}, and so we anticipated observing the formation of a travelling wave solution in the azimuthal direction with the front localised far from the boundary at $\theta = - \pi / 2$. The top vertical edge $\Gamma$ (see Fig. \ref{fig:solution_profiles}\textbf{a}), specified by $\big \{(r, \theta) \, | \, \theta = \pi / 2, r \in [ r_0 - \delta /2 , r_0 + \delta /2  ] \big \}$, was supplied with the Dirichlet boundary condition $u(r, \pi / 2, t)= 1$; von-Neuman no-flux boundary conditions were imposed on the three other edges. An initial condition $u(r, \theta, 0) = 0$ for  $(r, \theta) \in \Omega/\Gamma$ was used. For comparison, we performed equivalent simulations on a rectangle of width $\delta$ and length $\pi r_0$. We imposed the Dirichlet boundary condition $u(0, y, t) = 1$ on the edge $\Gamma '$ specified by $\big \{ (x, y) \, | \, x=0, y \in [0, \delta] \big \}$. 

Fig. \ref{fig:solution_profiles}\textbf{a} shows the solutions on the annulus and rectangle, which are qualitatively similar. The close qualitative match of our solutions, which for longer times appeared as travelling waves in the $x$ (rectangle) or $\theta$ (annulus) coordinates, encouraged us to examine the solution on the annulus for small times in Fig. \ref{fig:solution_profiles}\textbf{b} with a radially varying initial condition on $\Omega$. Taking $u(x, y, 0) = \exp{ \left ( \left ( x^2+(y-1)^2 \right ) / \alpha^2 \right ) },$ with $\alpha^2=0.05$, we found that the radial dynamics dominate the solution at early times. The solution approaches the stable steady state $u^*=1$ in the radial direction, and over a longer timescale begins to propagate in the azimuthal direction with little variation of $u$ in the radial coordinate.

\subsection{Differences between the solution on the rectangle and annulus}
\begin{figure}[t!]
    \centering
    \includegraphics{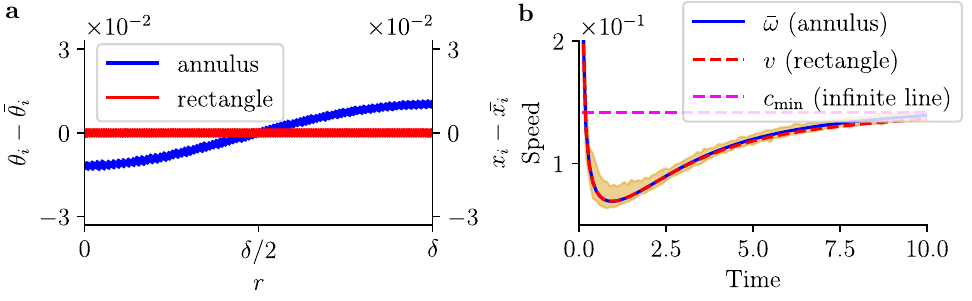}
    \caption{Properties of the $u = 0.5$ isoline in equivalent simulations on the rectangle and annulus at different times. \textbf{a} Deviation of $\theta_i$ of the isoline positions from the mean angle $\bar{\theta_i}$ in simulations on the annulus, and of the $x_i$ from the mean $x$-coordinate $\bar{x}$ on the rectangle. Both isolines are displayed at $t=2.0$. \textbf{b} Values of the mean angular speed $\bar{\omega} = \langle \omega_i(t) \rangle$ of the points on the isoline on the annulus  (blue) and linear speed $v = \langle v_i(t) \rangle$ on the annulus. The range of values of $\omega_i(t)$ for the annulus are shown (gold overlay). The minimum speed for a travelling wave solution allowed by Fisher's equation on an infinite line is also shown (magenta).}
    \label{fig:isoline_rect_annulus}
\end{figure}

Despite being qualitatively similar, the solution on the annulus does not perfectly match the solution on the rectangle.  To examine to what extent the solution deviates, we examined the coordinates $(x_i, y_i)$ and $(r_i, \theta_i)$ on the $u = 0.5$ isoline on the rectangle and the annulus, respectively. We plotted the angles $\theta_{i}$ (annulus) or $x$-coordinates $x_{i}$ (rectangle) for points on the isoline (Fig. \ref{fig:isoline_rect_annulus}\textbf{a}). On the rectangle, we found as expected that the isoline lies on a line of constant $x$ at any given time, with the solution propagating in the $x$-direction. For the annulus, the solution propagates azimuthally, similar to the rectangle, but points on the isoline deviated from the mean  $\bar{\theta_i}$ of the $\theta_{i}$, with the points at the inner radius at smaller angles than points at the outer radius. Interestingly, we saw the variation in $\theta_{i}$ was preserved at later times.

We also examined the angular speed $\omega_i$ (annulus) and linear speed $v_i$ (rectangle) of the points on the isoline (Fig. \ref{fig:isoline_rect_annulus}\textbf{b}), by calculating their velocity as in \cite{Buenzli2022}, giving
\begin{equation*}
    \mathbf{v}_i(x_i, y_i) = \frac{\partial u}{\partial t}\frac{1}{|\nabla u |^2} \nabla u = \mathbf{v}_i(r_i, \theta_i) \implies  \omega_i = \frac{\mathbf{v}_i(r_i, \theta_i) \cdot \mathbf{\hat{\theta}}_i}{r_i},
\end{equation*}
with $ \mathbf{\hat{\theta_i}} = \mathbf{\hat{x}}\cos{\theta_i} - \mathbf{\hat{y}}\sin{\theta_i}$ the usual polar azimuthal vector. The mean speed of both isolines is similar and for later times, both mean speeds approach the theoretical speed for the Fisher-KPP equation in one dimension. On both the rectangle and annulus, it is important to stress we have not observed the constant speed, fixed profile travelling waves typically used in analysis of the Fisher-KPP equation as we are not in the $t \to \infty$ limit.

\subsection{Annulus width and radius determine size of deviation of solutions.}
\begin{figure}[t!]
    \centering
    \includegraphics{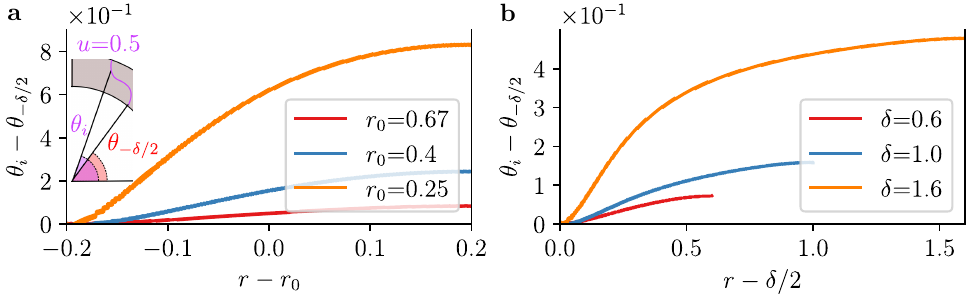}
    \caption{Profiles for the $u = 0.5$ isoline at $t=4$ for varying radius $r_0$ or thickness $\delta$. \textbf{a} Simulations utilising different values of $r_0=0.25, 0.4, 0.67$ ($\epsilon=1.6, 1, 0.6$) with fixed thickness $\delta = 0.4$. The difference between $\theta_i$ and the angle of the $u = 0.5$ isoline at the inner radius $r_0 - \delta /2$, $\theta_{- \delta/2}$ over the radial coordinate at $t=3.0$ is shown. The inset illustrates the definition of $\theta_{-\delta/2}$. \textbf{b} Simulations utilising different values of $\delta=0.6, 1.0, 1.6$ with fixed radius $r_0=1$ ($\epsilon=1.6, 1, 0.6$). The difference between $\theta_i$ and the angle of the $u = 0.5$ isoline at the inner radius $(r_0 - \delta /2)$, $\theta_{- \delta/2}$, over the radial coordinate at $t=3.0$, is shown.}
    \label{fig:midline-thickness}
\end{figure}

As we showed in Sec. \ref{sec:asymptotics}, radial gradients in the Fisher-KPP equation are on the length scale $\delta$ (annulus thickness), whereas derivatives in $\theta$ are coupled to $r$ and thus are related to both $r_0$ (annulus radius) and $\delta$. To characterise the range of $\epsilon$ that may be considered small, we simulated the Fisher-KPP equation with the same initial and boundary conditions as in Sec. \ref{sec:initial-numerics} but varying either $r_0$ or $\delta$ independently (Fig. \ref{fig:midline-thickness}). For all cases, the $\theta_i$ lagged behind $\theta_{-\delta / 2}$ (the angle of the $u=0.5$ isoline at the inner radius of the annulus). For annuluses with decreasing $r_0$ but constant $\delta$, there was an increased spread in the angles of the isoline. Increasing $\delta$ with $r_0$ constant showed an increasing spread in the $\theta_i$ from the inner to outer edge of the annulus. Varying both $\delta$ and $r_0$ between simulations by the same factor, however, showed only a small difference in the $\theta_i$ between simulations (not shown). This provides further evidence that $\epsilon$ governs the magnitude of azimuthal variations in the solution. In fact, we found that even for $\epsilon \sim 0.6$, the solution on the annulus is still close to forming azimuthal wavefronts with lines of constant $\theta_i$, analogous to the rectangle.

\section{Discussion}
We have demonstrated that the solutions yielded by the Fisher-KPP equation on annuluses with a small ratio of thickness to radius do not vary substantially to those on the rectangle. This is counter-intuitive; naively, we might expect that how the solutions differ from the rectangle might be described only by the radius $r_0$, which affects the tightness of the bend the solution propagates around. However, considering $\delta \to 0$ returns a one-dimensional Fisher-KPP equation only in $\theta$, in which there is trivially no variation in $r$. The small $\epsilon$ limit thus arises when the annulus approaches a circular arc. We further showed that the Fisher-KPP equation can be described to lowest order in such geometries by considering only a radial Fisher-KPP equation in $u$, and the azimuthal behaviour becomes important when the radial dynamics are at steady state.

Our results imply that on annuluses of sufficiently small thickness or large radius, it is justifiable to replace this geometry with a rectangle. The result extends straightforwardly to situations with more complex autonomous reaction terms than the logistic growth term present in the Fisher-KPP equation. However, a major limitation of the result is that it relies on the coupling of $r$ to $\theta$ in the Laplacian, and so can break down for non-autonomous reaction-diffusion equations. Thus, it will be necessary to extend or modify the result in such cases. Our findings have important implications for simulation of invasion on relevant natural geometries. In particular, we see that we can neglect the effect of curvature for a range of situations that may be relevant, for example, in the cranial neural crest, collective cell migration through torturous microchannels \cite{mazalan_effect_2020}, or reactions in annular geometries.

\pagebreak

\section*{Code availability}
\noindent A python implementation of the simulations described in the manuscript is available at \href{https://github.com/FusionLocus/fisher-kpp-annulus}{https://github.com/FusionLocus/fisher-kpp-annulus}.

\section*{Acknowledgements}
\noindent We thank members of the Baker Group for their helpful comments and input.

\section*{Author Contributions} 
\noindent P.K.M. \& R.E.B. conceived and supervised the study. J.J.P. contributed to study conceptualisation, performed the simulations and analysed the results. J.J.P., P.K.M. \& R.E.B. wrote and revised the manuscript.

\section*{Funding Details} 
\noindent J.J.P. is supported by funding from the Biotechnology and Biological Sciences Research Council (UKRI-BBSRC) [grant number BB/T008784/1]. This work is also supported by a grant from the Simons Foundation (MP-SIP-00001828, REB).

\noindent%

\bibliography{refs-2024-03-28}

\end{document}